# Entropy Production, Fractals, and Relaxation to Equilibrium

T. Gilbert* and J. R. Dorfman

*Department of Physics and Institute for Physical Science and Technology, University of Maryland, College Park, Maryland 20742*

P. Gaspard

*Center for Nonlinear Phenomena and Computer Systems, Université Libre de Bruxelles,
Code Postal 231, Campus Plaine, Blvd du Triomphe, B-1050 Brussels, Belgium*

(Received 7 March 2000)

The theory of entropy production in nonequilibrium, Hamiltonian systems, previously described for steady states using partitions of phase space, is here extended to time dependent systems relaxing to equilibrium. We illustrate the main ideas by using a simple multibaker model, with some nonequilibrium initial state, and we study its progress toward equilibrium. The central results are (i) the entropy production is governed by an underlying, exponentially decaying fractal structure in phase space, (ii) the rate of entropy production is largely independent of the scale of resolution used in the partitions, and (iii) the rate of entropy production is in agreement with the predictions of nonequilibrium thermodynamics.

A central issue in the statistical mechanics of nonequilibrium processes in fluids is to understand the irreversible production of entropy. Significant progress has been made recently in the theory for entropy production in nonequilibrium, Hamiltonian systems, using ideas from dynamical systems theory. Gaspard [1], Tél, Vollmer, and Breymann [2–5], Gilbert and Dorfman [6], as well as Tasaki and Gaspard [7,8], have described coarse graining procedures applied to the Gibbs entropy, defined in the phase space of systems maintained in nonequilibrium steady states. These procedures lead to results in accord with the predictions of the thermodynamics of irreversible processes. Although most of this work has been illustrated for a very simple system, a multibaker map, we believe that the main ideas and results can be generalized to more complicated many-body systems, even taking into account, of course, the more complex dynamics of such systems. Central to the results obtained so far is the fact that, in nonequilibrium steady states, and in the thermodynamic limit, the entropy production is controlled by fractal structures in the phase space distribution function. The presence of these fractal structures in the stationary state measures is crucial for the theory of entropy production, since if the distribution functions were smooth, the usual Gibbs entropy arguments would apply, and there would be no change in the Gibbs entropy and no positive irreversible entropy production. This previous work left open the question as to how one might justify the irreversible entropy production for systems freely relaxing to a uniform equilibrium state, based upon the existence of fractal structures underlying the relaxation process.

The purpose of this Letter is to show that one can also understand entropy production in the approach to equilibrium, in a fashion similar to that in a nonequilibrium steady state. That is, fractal structures appear for systems approaching an equilibrium state, and the treatment of the entropy production, following the lines initiated by Gaspard, leads to the well-known results of irreversible thermodynamics [9]. Thus, irreversible entropy production in time dependent processes as well as in nonequilibrium steady states can be understood from a single point of view and treated by closely related, essentially identical, analytical methods.

The procedure we follow uses the contribution to the phase space distribution of the *singular* hydrodynamic modes of diffusion of the Frobenius-Perron operator, as described by Gaspard [10,11]. The contribution of these hydrodynamic modes to the irreversible entropy production is determined by using a procedure based upon a partitioning of the phase space into small regions, which allows the fractal structures in the phase space distribution functions to be described. This method naturally raises the issue of what size and what type of partitions should be used in this process [12]. Here we use a natural Markov partition, based upon the chaotic dynamics of the system, and we show that the physical results are essentially independent of the size of the partitions used, provided the partitions are not larger than some characteristic microscopic size of the system. We will illustrate the ideas by means of a multibaker system relaxing to equilibrium, which is easy to describe. In further work we will show how these methods can be extended to treat diffusion in a periodic Lorentz gas [13], and hydrodynamic relaxation to equilibrium in fluid systems [14].

We consider the relaxation to an equilibrium distribution for a large ensemble of random walkers on sites labeled by integers on a chain of length $L$, with periodic boundary conditions. The ensemble is such that the initial distribution of walkers is nonuniform on the $L$ sites. Furthermore, to avoid some mild complications due to the lack of irreducibility of the corresponding transition matrix of the simplest binary process where particles hop to the right or left, we consider here a process where the random walkers have equal probabilities of hopping to the right

   0031-9007/00/85(8)/1606(4)$15.00   © 2000 The American Physical Society

or left or of remaining at their position. We now replace this simple random process by a deterministic one that mimics the Hamiltonian dynamics underlying deterministic diffusion. Therefore we use a triadic multibaker map to describe the system with a deterministic, reversible, and area-preserving time evolution. The phase space dynamics take place on $\mathcal{L} \times [0,1]^2$, where $\mathcal{L} = \{1, \ldots, L\} \subset \mathbf{Z}$. The time evolution is given by

$$M(n,x,y) = \begin{cases} (n-1, 3x, \frac{y}{3}), & 0 \le x < \frac{1}{3}, \\ (n, 3x-1, \frac{y+1}{3}), & \frac{1}{3} \le x < \frac{2}{3}, \\ (n+1, 3x-2, \frac{y+2}{3}), & \frac{2}{3} \le x < 1. \end{cases} \quad (1)$$

Here the integer index $n$ denotes the particular site on the chain, and $x, y$ are internal coordinates used to make the motion deterministic. Since $(x, y) \in [0, 1]^2$, each site of the chain is a square. We use periodic boundary conditions to identify site $j$ with site $L + j$, etc.

Let $\rho_t(n, x, y)$ be the time dependent distribution function for particles on the phase space. Using the dynamics given in Eq. (1), we see that $\rho_t(n, x, y)$ satisfies the Frobenius-Perron equation $\rho_{t+1}(n, x, y) = \rho_t[M^{-1}(n, x, y)]$. The mean number of particles in a cell $B$ of the $n$th square $A_n$ is given by $\mu_t(B) = \int_B dx\, dy\, \rho_t(n, x, y)$, and this number evolves in time according to $\mu_{t+1}(B) = \mu_t(M^{-1}B)$.

To display the fractal form that underlies the relaxation to equilibrium and the corresponding entropy production, we define the cumulative function $g_t(n, y) = \int_0^1 dx \int_0^y dy'\, \rho_t(n, x, y')$, with time evolution determined by the Frobenius-Perron operator [15,16]:

$$g_{t+1}(n,y) = \begin{cases} \frac{1}{3} g_t(n+1, 3y), & 0 \le y < \frac{1}{3}, \\ \frac{1}{3} g_t(n, 3y-1) + \frac{1}{3} g_t(n+1, 1), & \frac{1}{3} \le y < \frac{2}{3}, \\ \frac{1}{3} g_t(n-1, 3y-2) + \frac{1}{3} g_t(n+1, 1) + \frac{1}{3} g_t(n,1), & \frac{2}{3} \le y < 1, \end{cases} \quad (2)$$

where we have assumed that the distribution function $\rho_t(n, x, y)$ is uniform with respect to $x$. This follows from the fact that the $x$ direction corresponds to the expanding direction for the dynamics, and the distribution *within each cell* will become uniform with respect to $x$ on an exponentially short time scale compared with the time necessary for the total distribution to become uniform over the whole chain of length $L$. For the particular value $y = 1$, we note that $g_t(n, 1)$ satisfies

$$g_{t+1}(n, 1) = \frac{1}{3} g_t(n+1, 1) + \frac{1}{3} g_t(n, 1) + \frac{1}{3} g_t(n-1, 1), \quad (3)$$

which is a simple random walk equation for a particle at integer points on a line (more properly, a circle), corresponding to probabilities of $1/3$ each, for moving to the left, to the right, or staying in place, at each time step. This is a diffusive process with diffusion coefficient, $D = 1/3$.

When the system has entirely relaxed to equilibrium, the cumulative measure has the asymptotic form $g_\infty(n, y) = \rho_{\text{eq}} y$, where $\rho_{\text{eq}}$ is the uniform equilibrium density. In order to study the details of the relaxation process, it is natural to write the cumulative function $g_t$ as a superposition of the eigenmodes $\psi_k(n) = \exp(ikn)$ of Eq. (3), where the wave number $k$ takes values restricted to $2\pi m/L$ with $m \in \mathbf{Z}$ modulo $L$, by the periodic boundary conditions. That is, we write

$$g_t(n, y) = \sum_k \chi_k^t a_k \psi_k(n) F_k(y), \quad (4)$$

where $\chi_k = (1/3)(1 + 2\cos k)$ is the eigenvalue corresponding to the eigenvector $\psi_k(n)$ and the coefficients $a_k$ are determined by the initial distribution. Since $g_t(n, y)$ is real, the coefficients $a_k$ satisfy the property that $a_k^* = a_{-k}$. The equilibrium solution corresponds to $k = 0$, i.e., $a_0 = \rho_{\text{eq}}$. All of the modes with $k \ne 0$ and $\pi$ have a double degeneracy. In particular, the slowest decaying modes correspond to $k = \pm 2\pi/L$. In what follows, we will be concerned with the slowest decaying modes, and restrict our attention to these particular values of $k$.

From Eq. (2), the functions $F_k(y)$ are found to satisfy the functional equation

$$F_k(y) = \begin{cases} \frac{\exp(ik)}{3\chi_k} F_k(3y), & 0 \le y < \frac{1}{3}, \\ \frac{1}{3\chi_k} F_k(3y-1) + \frac{\exp(ik)}{3\chi_k} F_k(1), & \frac{1}{3} \le y < \frac{2}{3}, \\ \frac{\exp(-ik)}{3\chi_k} F_k(3y-2) + \frac{\exp(ik)+1}{3\chi_k} F_k(1), & \frac{2}{3} \le y < 1. \end{cases} \quad (5)$$

This is a de Rham–type functional equation [17], the solutions of which can be constructed iteratively, and converge rapidly, compared to hydrodynamic relaxation times, to the cumulative function of the hydrodynamic eigendistribution of the Perron-Frobenius operator. These eigendistributions are fractal functions of the $y$ coordinate. The correspondence between these cumulative functions and the hydrodynamic modes of the Perron-Frobenius operator can be made more explicit if, for small $k$ or, equivalently, large $L$, we expand $F_k$ in powers of $k$, to obtain $F_k(y) = y + ikT(y) + O(k^2)$, which corresponds to a gradient expansion. The lowest order term is just the cumulative function for the uniform system, while the function $T$ in the first order term is the triadic equivalent of the dyadic Takagi function already familiar in the nonequilibrium eigenmodes of the multibaker map [15]. Here $T(y)$ satisfies the functional equation

$$T(y) = \begin{cases} \frac{1}{3}T(3y) + y, & 0 \le y < \frac{1}{3}, \\ \frac{1}{3}T(3y-1) + \frac{1}{3}, & \frac{1}{3} \le y < \frac{2}{3}, \\ \frac{1}{3}T(3y-2) + 1 - y, & \frac{2}{3} \le y < 1. \end{cases} \quad (6)$$

This functional equation has a unique solution which is a continuous but nondifferentiable function depicted in Fig. 1.

This function has the following remarkable fractal properties. If we consider a partition of the unit square in $3^{d+1}$ horizontal cells denoted by the sequence $\underline{\omega}_{d+1} \equiv \omega_0 \omega_1 \omega_2 \cdots \omega_d$ with $\omega_i \in \{0, 1, 2\}$ for all $i$, the end points of those cells are the set of points with the corresponding triadic expansion,

$$y(\underline{\omega}_{d+1}) = \sum_{i=0}^{d} \frac{\omega_i}{3^{i+1}}, \quad (7)$$

for all possible sequences $\underline{\omega}_{d+1}$. The difference $\Delta T(\underline{\omega}_{d+1})$ between the Takagi function $T(y)$ evaluated at the two end points of the corresponding cell has the scaling property

$$\Delta T(\underline{\omega}_{d+1}) = \frac{1}{3} \Delta T(\omega_1 \omega_2 \cdots \omega_d) + \frac{1 - \omega_0}{3^{d+1}}, \quad (8)$$

which is a consequence of Eq. (6). This relation shows that the Takagi function is self-affine because the function in the larger cell $\omega_1 \omega_2 \cdots \omega_d$ is mapped onto the function in the smaller cell $\underline{\omega}_{d+1}$ by a scaling transformation. The connection between $T(y)$ and the more familiar hydrodynamic modes of kinetic theory or of linearized hydrodynamics is clarified by the observation that the derivative of $T(y)$ with respect to $y$ is the total displacement of a particle that starts at this particular value of $y$ at the initial time. This displacement is, in this discrete time map, the analog of the time integral of the velocity of a particle undergoing deterministic diffusion in a fluid system, and is a wildly varying function of $y$. The function $T(y)$ itself, while fractal, is much more tractable than its derivative [15,16]. In this way it is possible to establish the connection between the approach here, using the exact dynamics, and the Green-Kubo formulas for transport coefficients and thus indicate how results obtained here may be applied to many particle systems [10,11].

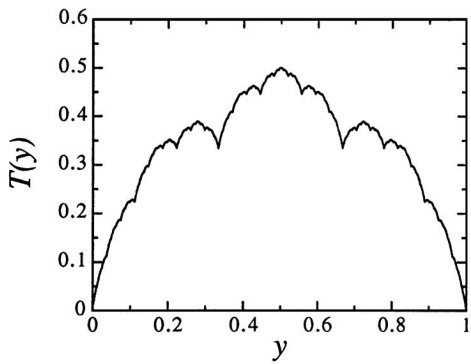

FIG. 1. The triadic Takagi function constructed iteratively from Eq. (6).

Using the symmetry between the modes $k = \pm 2\pi/L$, we can express $g$ as

$$g_t(n, y) = \rho_{\text{eq}} y + \chi_k^t [2 \operatorname{Re}(a_k e^{ikn}) y \\ - 2k \operatorname{Im}(a_k e^{ikn}) T(y) \\ + O(k^2)]. \quad (9)$$

We now show how the hydrodynamic mode analysis relates to the irreversible production of entropy during the final stages of the approach to a uniform equilibrium state. First we consider the macroscopic description. According to the usual arguments [1,9], the phenomenological entropy production in a one-dimensional diffusive system with configuration-space density function is given by

$$\sigma_t(r) = \frac{D}{c_t(r)} \left[ \frac{\partial c_t(r)}{\partial r} \right]^2, \quad (10)$$

where $c_t(r) = g_t(n, 1)$ is the concentration of particles at the position $r$ if we take the continuous limit and replace $n$ by a continuous variable $r$. In our system, we can calculate the concentration using Eq. (9) and the fact that $T(y = 1) = 0$, which is a consequence of Eq. (6). Accordingly, the entropy production (10) becomes

$$\sigma_t(r) = \frac{\chi_k^{2t}}{3\rho_{\text{eq}}} [2k \operatorname{Im}(a_k e^{ikr}) + O(k^2)]^2, \quad (11)$$

an expression to which we will return shortly.

As described by Gaspard [1,11] and Gilbert and Dorfman [6], the fractal structures are responsible for the positivity of the rate of entropy production for a hyperbolic or Anosov-like dynamical system. The time dependent entropy production can be introduced by first considering the coarse grained entropy of the particle distribution in the square $A_n$ partitioned into cells $\{B\}$,

$$S_t(A_n, B) = \sum_{B \subset A_n} \mu_t(B) \ln \frac{\nu(B)}{\mu_t(B)}, \quad (12)$$

where $\mu_t(B)$ is the mean number of particles in the cell $B$ and $\nu(B)$ is its area. The time change of this entropy, $\Delta S_t = S_{t+1} - S_t = \Delta_e S_t + \Delta_i S_t$, has a contribution from the entropy flux between the square $A_n$ and its neighbors, $\Delta_e S_t(A_n, B) = S_t(M^{-1}A_n, B) - S_t(A_n, B)$, and another from the entropy production inside $A_n$ itself, $\Delta_i S_t(A_n, B) = S_{t+1}(A_n, B) - S_{t+1}(A_n, MB)$, which is inferred from the previous relations and $S_t(M^{-1}A_n, B) = S_{t+1}(A_n, MB)$. If we partition the square $A_n$ into cells $B$ such that their images $MB$ are the aforementioned cells denoted by the sequence $\underline{\omega}_{d+1}$, the time dependent entropy production rate is given by

$$\Delta_i S_t(A_n, B) = \sum_{\underline{\omega}_{d+1}} \Delta g_{t+1}(n, \underline{\omega}_{d+1}) \ln \frac{3\Delta g_{t+1}(n, \underline{\omega}_{d+1})}{\Delta g_{t+1}(n, \underline{\omega}_d)}. \quad (13)$$

By using Eq. (9), we can write the measure of the horizontal cell $MB$ corresponding to $\underline{\omega}_{d+1}$ as

$$\Delta g_{t+1}(n, \underline{\omega}_{d+1}) = \frac{\rho_{\text{eq}}}{3^{d+1}} + \chi_k^t \left[ 2 \operatorname{Re}(a_k e^{ikn}) \frac{1}{3^{d+1}} - 2k \operatorname{Im}(a_k e^{ikn}) \Delta T(\underline{\omega}_{d+1}) + O(k^2) \right], \quad (14)$$

with the difference $\Delta T(\underline{\omega}_{d+1})$ given by Eq. (8).

Taking $\chi_k^t$ to be a small parameter in the limit $t \to \infty$, we can expand the entropy production, Eq. (13), about its vanishing equilibrium value, and keep terms up to quadratic in the deviation from equilibrium. We find, after a few manipulations, that the entropy production rate reduces to

$$\Delta_i S_t(A_n, B) \simeq \frac{\chi_k^{2t}}{2\rho_{\text{eq}}} [2k \operatorname{Im}(a_k e^{ikn}) + O(k^2)]^2 \\ \times \left[ 3^{d+1} \sum_{\underline{\omega}_{d+1}} \Delta T(\underline{\omega}_{d+1})^2 \right. \\ \left. - 3^d \sum_{\underline{\omega}_d} \Delta T(\underline{\omega}_d)^2 \right]. \quad (15)$$

By using the fractal property (8) of the triadic Takagi function, the term in brackets in the above expression can be reduced to

$$3^{d+1} \sum_{\underline{\omega}_{d+1}} \Delta T(\underline{\omega}_{d+1})^2 - 3^d \sum_{\underline{\omega}_d} \Delta T(\underline{\omega}_d)^2 = \frac{2}{3}. \quad (16)$$

As a consequence, the theoretical entropy production (15) becomes equal to the phenomenological entropy production (11). We have therefore shown that, in the limit of large $L$ where we can neglect the $O(k^2)$ terms in the gradient expansion, the entropy production formula [Eq. (13)] yields a result in full agreement with the phenomenological prescriptions of nonequilibrium thermodynamics.

In conclusion, we note that for this simple model we have shown that even during the time that the system is relaxing to equilibrium, and especially in the final stages of the relaxation, the nonequilibrium distribution function has fractal properties and that these properties are responsible for the positivity of the irreversible entropy production. Elsewhere [1,6,11] we have argued that, if nonequilibrium distributions were smooth and differentiable, the Gibbs entropy would not show any irreversible behavior. These arguments apply to this case also and strongly suggest that the positivity of irreversible entropy production for Hamiltonian systems is due to the singularity of the phase space distributions, for systems relaxing to equilibrium as well as for systems in nonequilibrium steady states maintained by reservoirs. Further we note that the rate of entropy production in the Navier-Stokes regime, $O(1/L^2)$, at least, is independent of the coarse graining parameter $d$. We would hope, though it remains to be seen, that this independence is a general feature of irreversible entropy production.

We thank E. G. D. Cohen, R. Klages, G. Nicolis, L. Rondoni, S. Tasaki, T. Tél, and J. Vollmer for discussions. J. R. D. thanks the National Science Foundation for support under Grant No. PHY 98-20824. P. G. thanks the FNRS Belgium and the IAP Program of the Belgian Federal OSTC for financial support.